\documentclass[aps,prl,amsmath,amssymb,twocolumn,superscriptaddress]{revtex4}
\usepackage{bm}
\usepackage{rotating}
\usepackage{aas_macros}

\newcommand\lsim{\mathrel{\rlap{\lower4pt\hbox{\hskip1pt$\sim$}}
        \raise1pt\hbox{$<$}}}
\newcommand\gsim{\mathrel{\rlap{\lower4pt\hbox{\hskip1pt$\sim$}}
        \raise1pt\hbox{$>$}}}
\newcommand\propsim{\mathrel{\rlap{\lower4pt\hbox{\hskip1pt$\sim$}}
        \raise1pt\hbox{$\propto$}}}
\newcommand{\D}{\mathrm{d}}

\newcommand{\ret}{\mathrm{ret}}
\newcommand{\Msun}{\mathrm{M}_{\odot}}

\newcommand{\GW}{\mathrm{GW}}

\newcommand{\Edd}{\mathrm{E}}
\newcommand{\Iota}{Y}
\newcommand{\cool}{\mathrm{c}}
\newcommand{\q}{q_0}

\begin{document}
\bibliographystyle{apsrev}

\title{Brightening of an Accretion Disk Due to Viscous Dissipation of\\
Gravitational Waves During the Coalescence of Supermassive Black Holes}

\author{Bence Kocsis}
\affiliation{Harvard-Smithsonian Center for Astrophysics, 60 Garden Street,
Cambridge, MA 02138, USA}

\author{Abraham Loeb}
\affiliation{Harvard-Smithsonian Center for Astrophysics, 60 Garden Street,
Cambridge, MA 02138, USA}

\date{\today}

\begin{abstract}
Mergers of supermassive black hole binaries release peak power of up to
$\sim 10^{57}~{\rm erg~s^{-1}}$ in gravitational waves (GWs). As the GWs
propagate through ambient gas, they induce shear and a small fraction of
their power is dissipated through viscosity.  The dissipated heat appears
as electromagnetic (EM) radiation, providing a prompt EM counterpart to the
GW signal.  For thin accretion disks, the GW heating rate exceeds the
accretion power at distances farther than $\sim 10^3$ Schwarzschild radii,
independently of the accretion rate and viscosity coefficient.
\end{abstract}

\maketitle

\paragraph*{Introduction.---}
Coalescing binaries of supermassive black holes (SMBHs) are the primary
sources of gravitational waves (GWs) for the planned Laser Interferometric
Space Antenna (LISA
\footnote{http://lisa.nasa.gov/}). Recent advances in numerical relativity
enable to forecast the GW luminosity and waveform during the final phase of
a SMBH coalescence event (e.g. Ref.~\cite{Baker07,Berti07,Buonanno07} and
references therein). The peak GW luminosity , $L_{\GW}\sim 10^{56-57}~{\rm
erg}~{\rm s}^{-1}$, will be observable with LISA out to high cosmological
redshifts.

As the GWs propagate away from their source, they interact with matter in
several ways. The shear they induce in surrounding gas can be dissipated
through viscosity \cite{Hawking66}.  The GWs could also drive transverse
and longitudinal density waves \cite{Esposito71}, excite resonant vibration
modes, boost the frequency of photons \cite{Marklund00,bms01}, lead to
graviton--photon conversion \cite{Cuesta02,Kallberg06}, and couple to
Alfven and magnetohydrodynamic waves in strongly magnetized plasmas
\cite{mn83,Papadopoulos01,Clarkson04,Kallberg04,Forsberg08}. These
interactions are so weak that GWs are expected to escape from the densest
environments like the cores of supernova explosions, gamma--ray bursts, or
the early universe, and travel across cosmological distances without any
noticeable attenuation. However, in the vicinity of coalescing SMBH
binaries, even a miniscule coupling with matter could lead to a bright
electromagnetic (EM) signal. In this {\it Letter}, we demonstrate that the
viscous dissipation of GWs in the astrophysical environments of SMBH
binaries might be detectable.

The merger dynamics of a pair of gas-rich galaxies with SMBHs generically
channels large quantities of gas to the central region and creates a
gaseous envelope around the resulting SMBH binary \cite{Springel}.  The
presence of gas and stars is expected to catalyze the hardening of the
SMBH binary \cite{Escala05,Dotti07} to a separation where GW emission --
on its own -- would be capable of shrinking the orbit on a Hubble time
\cite{BBR}. The high infall rate of gas in this environment is expected to
lead to the formation of a geometrically-thin accretion disk \cite{Escala05}.
For binaries of nearly equal mass SMBHs, the tidal field of the binary would open a
central cavity in the disk with a radius of about twice the orbital radius
of the binary \cite{mp05}. During the final phase of SMBH coalescence, the
cavity would not be able to track the ever increasing rate at which GW
emission shrinks the binary orbit and so the cavity radius would freeze at
$r_{\min}\sim 120 r_S \q^{0.45} M_7^{0.07}$. Here, $r_S=2{\rm G}M/{\rm
c}^2$, $\q=4q/(1+q)^2$ is the scaled symmetric mass ratio, and
$M_7=M/(10^7\Msun)$ is the total binary mass.  The GW dissipation in this
punctured disk should inevitably lead to an EM transient.

If the prompt EM counterpart to a GW signal is sufficiently bright, it
would enable observations of SMBH mergers with traditional telescopes long
before LISA becomes operational.
Its detection would provide a unique test of general relativity in the
strong--field regime. A successful identification of an EM counterpart to a
LISA source would have far--reaching consequences \cite{Kocsis07}.

Recent studies considered other mechanisms for EM counterparts to SMBH
mergers: {\it (i)} periodic variation in the gravitational potential owing
to the orbital motion of the binary during the early stages of the inspiral
\cite{mm08}; {\it (ii)} shocks induced by the sudden mass loss of the
binary due the final GW burst \cite{bp07}; {\it (iii)} shocks induced by a
supersonic gravitational recoil kick \cite{Lippai08,sk08,sb08}; and {\it
(iv)} infall of gas onto the SMBH remnant \cite{mp05}.
We expect the viscous GW heating effect to dominate over mechanism {\it
(i)} during the late inspiral at distances much larger than the binary
separation, since the tidal potential driving effect {\it (i)} scales as
$d^2/r^3$.  The viscous dissipation of GWs is unique in its ability to
yield a {\it prompt} EM counterpart within hours--days after the peak GW
burst, because it is driven by the time-evolution of the GW luminosity.  In
contrast, the mass loss effect of mechanism {\it (ii)} requires the orbital
timescale of days--weeks necessary for the gas to respond and shocks to
build up \cite{sk08}.  The supersonic kick from mechanism {\it (iii)}
becomes effective after the disturbance to the disk propagates out to the
radius where it is supersonic, over a timescale of months--years
\cite{Lippai08,sk08,sb08}. Finally, infall of gas in mechanism {\it (iv)}
occurs only years after coalescence \cite{mp05}.

\paragraph*{Gravitational waves from a black hole merger.---}
The GWs produced by a black hole binary merger are dominated by the spin-2
weighted $l=2$, $m=\pm 2$ spherical tensor harmonic
\cite{Berti07,Buonanno07}. We approximate the GW energy flux by
its asymptotic behavior at large distances \cite{Isaacson68}. In spherical coordinates,
\begin{equation}\label{e:eGW}
 e_{\GW}(t,r,\theta,\phi)= \Iota(\theta) \frac{L_{\GW}(t_{\ret})}{4\pi
 {\rm c}r^2},
\end{equation}
where $\theta$ is the angle relative to the total angular momentum vector,
$\Iota(\theta)=(5/2)[\sin^8(\theta/2)+\cos^8(\theta/2)]$ has a unit
average on the sphere, and $L_{\GW}(t_{\ret})$ is the GW luminosity at
retarded time $t_{\ret}=t-r/{\rm c}$,
which we approximate as $L_{\GW}(t)\propto \q^2(|\q t-t_1|/t_1)^{-5/4}$ for $t<0$, and
$\exp(-{\rm c}t/(2.5r_S))$ for $t>t_1$, while being constant in between. We set the overall
scaling and $t_1$ to match the Newtonian inspiral luminosity for $t\ll 0$,  and the
peak luminosity from numerical simulations
at $t=0$,  $L^{\GW}_{\rm insp}=32^{-1} (5/64)^{1/4} {\rm c}^{5}{\rm G}^{-1} \q^{3/4}$
and $L^{\GW}_{\rm peak}\approx 10^{-3}{\rm c}^{5}{\rm G}^{-1} \q^2$,
respectively (see Fig.~\ref{f:lightcurve} below). This is
correct within a factor of $\sim 2$ depending on the magnitude and orientation of
SMBH spins \cite{Berti07,Buonanno07}. The total GW energy released is
$\Delta E_{\GW} = \int_{-\infty}^{\infty} L_{\GW}\D t=\kappa M{\rm c}^2$,
where $3\% \lsim \kappa/\q^2\lsim 7\%$, and $5\%$ using our simple fit. The timescale
of the intense GW burst is $\Delta t_{\GW} = \kappa M {\rm c^2}/L^{\GW}_{\rm peak}
\sim 20 r_S/c$.

\paragraph*{GW dissipation in a viscous medium.---}
The stress-energy tensor of a viscous medium is augmented by $T_{\mu \nu} =
-2 \eta \sigma_{\mu \nu}$, where $\eta$ is the dynamical shear viscosity
coefficient and $\sigma_{\mu \nu}$ is the fluid's rate of shear \cite{Weinberg}.
As GWs traverse the fluid they induce a shear $\sigma_{\mu
\nu}= \frac{1}{2} \dot h_{\mu \nu}$
\footnote{Note that this expression is exact only for a
stationary homogeneous fluid in flat space,
as appropriate here at large radii.},
where an overdot denotes a time
derivative of the metric perturbation, $h_{\mu\nu}$. The weak-field
Einstein equation yields
$\Box h_{\mu \nu} = -16 \pi G\eta \dot h_{\mu \nu}/c^4$
\cite{Hawking66,Weinberg},
implying that the GW energy density $e_{\rm GW}$ is dissipated at the rate,
\begin{equation}\label{e:heat}
 \dot e_{\rm heat} \equiv -\dot e_{\rm GW}= \frac{16 \pi G \eta}{c^2}
 e_{\rm GW}.
\end{equation}
Thus, the energy density dissipates exponentially with a time constant of
$t_d = (16 \pi {\rm G}\eta/{\rm c}^{2})^{-1}$.

\paragraph*{Heating a Thin Accretion Disk.---}
In the standard ``$\alpha$--model'' of radiatively-efficient accretion
flows \cite{Shak,Thorne,Pringle81}, the gas orbits around the central SMBH
within a thin co-planar disk, characterized by a vertical height $H(r)\ll
r$ and a low temperature $T\lesssim 10^6~$K. The viscosity coefficient is
parameterized as $\eta(r) = (2/3) \alpha P(r)/\Omega(r)$
\footnote{The $\alpha$-viscosity results from
magneto-hydrodynamic turbulence, whose cascade dissipates energy from shear
motion. We assume that the shear caused by GWs is dissipated by the same
turbulence that transports angular momentum in the disk, according to
Eq.~(\ref{e:heat}).}, where $\Omega(r)\equiv {\rm G} M/r^{3}$ is the
angular velocity, $\alpha\lesssim 1$, and $P(r)$ is either the gas
pressure, $P_{\rm gas}$, or the total (gas$+$radiation) pressure, $P_{\rm
tot}$. We therefore write $P(r)=P_{\rm gas}^{b}P_{\rm tot}^{1-b}$ with
$0<b<1$. The mass accretion rate $\dot M$ can be expressed in terms of the
Eddington luminosity $L_{\Edd}(M)$ as $\dot M= \dot m \dot M_E$, where
$\dot M_E=\epsilon L_E/{\rm c^2}$ and $\epsilon$ is the fraction of the
accreted mass which gets converted into radiation. Once the disk opacity
$\kappa$ is specified, the accretion model is fixed by the free parameters
$\dot m$, $\epsilon$, $b$, and $M$ \cite{Goodman03,gt04}. The physical
characteristics of the gas, such as the midplane temperature $T(r)$,
surface mass density $\Sigma(r)$, and the scale-height $H(r)$, can be all
expressed in terms of these parameters.

The GW energy absorbed per unit area of the disk surface, $H \dot e_{\rm
heat}$, depends on the combination $\eta H$ in
Eq.~(\ref{e:heat}). Remarkably, this particular combination is simply
related to the mass accretion rate ${\dot M}$ in a steady state
\cite{Pringle81},
\begin{equation}\label{e:conservation}
 \eta(r)H(r) = \frac{\nu(r)\Sigma(r)}{2} = \frac{\dot M}{6\pi} = \rm const ,
\end{equation}
where, $\nu=\eta/\rho=2H \eta /\Sigma$ is the kinematic viscosity and
$\rho$ is the mass density of the gas.

Substituting Eqs.~(\ref{e:eGW}) and (\ref{e:conservation}) into
Eq.~(\ref{e:heat}), we find that the rate of GW dissipation per unit
surface area is independent of the disk viscosity or opacity,
\begin{equation}\label{e:L_heat}
 H  \dot e_{\rm heat}  =\frac{16\pi \rm G}{{\rm c}^2} \eta H e_{\rm GW} = \frac{8}{3} \frac{\rm G}{{\rm c}^{3}} \dot M Y(\theta) \frac{L_{\GW}(t_{\ret})}{4\pi r^2}.
\end{equation}
Thus, {\it equal amounts of heat are dissipated by the GWs per
logarithmical radius bin of the disk}.  The heating rate is
proportional to $\dot M$ but is otherwise independent of the disk
parameters.

The heating rate should be compared to the standard dissipation rate of the
accretion disk, $H \dot e_{\rm disk} = (3/8\pi){\rm G} M \dot M/r^3 $,
\begin{equation}\label{e:e_heat/e_disk}
 \frac{\dot e_{\rm heat}(t_{\ret},r)}{\dot e_{\rm disk}(r)} = \frac{32}{9}
 \Iota(\theta)\, r_3\, L^{\GW}_{-3}(t_{\ret}) ,
\end{equation}
where $L^{\GW}_{-3}= L_{\GW}/(10^{-3}{\rm c}^5/{\rm G})$. In general,
$L^{\GW}_{-3}\sim 1$ during the peak emission of GWs, independently of
$M$. Thus, the peak GW heating rate exceeds the standard disk dissipation
rate at $r_3\gtrsim 1$ by a factor of $(32/9) \Iota r_3$. If the disk
resides in the orbital plane of the binary, ${\dot e_{\rm heat}}/{\dot
e_{\rm disk}}=(10/9) r_3$.  {\it This result is universal and independent
of $M$ or the accretion disk parameters.}

\paragraph*{Cooling Timescale.---}
The excess heat deposited by GWs will eventually be radiated away
electromagnetically.  The EM light curve depends on the uncertain details
of the turbulent accretion disk, and in particular on the vertical
transport of heat.

\begin{figure}[tbh]
\centering
\mbox{\includegraphics[height=2.5in]{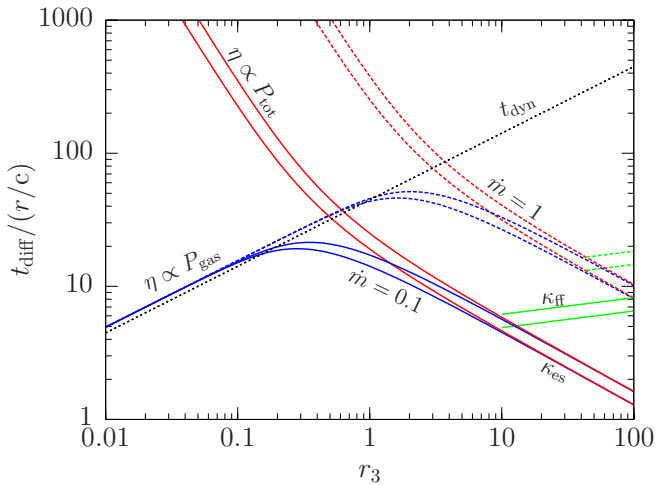}}
\caption{\label{f:tcool} The radiative diffusion time, $t_{\rm diff}$, over
the geometric light travel time, $r/{\rm c}$, as a function of radius for
various $\alpha$--models with $\alpha=0.3$.  The solid and dashed curves
correspond to $\dot m=0.1$ and $1$, respectively.
Two sets of curves are shown for $M_7=1$ and 10, respectively, with the
lower mass corresponding to the lower curves.
}
\end{figure}

For an optically-thick disk with a vertical optical depth $\tau(r)\gg1$,
the diffusion timescale of photons out of the midplane is $t_{\rm diff}
\sim \tau H/{\rm c}$ \cite{Rybicki,Goodman03}. The timescale for a patch of
the disk to change its thermal energy content by turbulent heat transport
is $t_{\rm therm} \sim t_{\rm dyn}/\alpha$, where $t_{\rm dyn}\sim H/c_s
\sim 1/\Omega$ and $c_s$ is the sound speed \cite{Krolik}. The excess
dissipation of heat by GWs will eventually be radiated away within a
timescale, $t_{\cool}=\min(t_{\rm diff},t_{\rm therm})$ \cite{gt04}\footnote{The EM
lightcurve  is dominated by emission from $0.1\lesssim r_3 \lesssim 1$ (Fig.~\ref{f:lightcurve}) for which
$t_{\rm diff}$ is similar to $t_{\rm therm}$ (Fig.~\ref{f:tcool}).}.

We calculate the diffusion timescale for various disk models, following
Refs.~\cite{Goodman03,gt04}. We consider a viscosity term that is
proportional to the gas or total pressure, an opacity coefficient that is
dominated by electron scattering with $\kappa_{\rm es}=0.35 {\rm cm}^2~{\rm
g}^{-1}$, or free-free transitions with $\kappa_{\rm ff}=6.9\times 10^{22}
\rho T^{-7/2}$~[cgs], and Eddington accretion rates of $\dot m=0.1$ and 1,
respectively. We also adopt $\alpha=0.3$ and $M_7=1$ or
10. Figure~\ref{f:tcool} shows the resulting $t_{\rm diff}(r)$. The disk
can be divided into three distinct regions: {\it (a)} an inner region
dominated by radiation pressure; {\it (b)} a middle region dominated by gas
pressure and $\kappa_{\rm es}$; and {\it (c)} an outer region dominated by
gas pressure and $\kappa_{\rm ff}$. In region {\it (a)}, the viscosity
prescription affects the results dramatically.  If $\eta\propto P_{\rm
gas}$ then $t_{\rm diff}\sim t_{\rm dyn}$, whereas if $\eta\propto P_{\rm
tot}$ then radiative diffusion is inefficient and $t_{\cool}\sim t_{\rm
therm}$. For typical parameters, the outer region corresponds to very large
radii, at which the disk may fragment due to its self-gravity.  We find
that for $\dot m=0.1$ and $0.1\lsim M_7\lsim 10$, the range of
\begin{equation}\label{e:tcool}
 3 \lsim {\rm c}t_{\cool}(r)/r \lsim 30,
\end{equation}
applies to all radii of interest. We note that the cooling time might be
reduced if the radiation escapes through regions of low gas density, if a
significant fraction of the turbulent magnetic energy is dissipated in
surface layers \cite{ms00}, or if a photon-bubble instability operates
\cite{Gammie98}.

\paragraph*{Brightening of the Disk.---}
For stationary disks, there is no radial heat transport on the relevant
timescales, since the viscous time is long, $t_{\cool}\ll t_{\rm
visc}\sim {r^2}/{\nu}\sim t_{\rm therm} (r/H)^2$
\cite{Pringle81,Krolik}. The flux excess due to GW heating, $\Delta
F(t,r)\equiv F(t,r)-F_{\rm disk}(t,r)$, is determined by an ordinary
first-order linear differential equation in time $t$,
\begin{equation}\label{e:heatingbal}
 t_{\cool} \Delta \dot F + \Delta F= H\dot e_{\rm heat},
\end{equation}
where the driving term is determined by the instantaneous GW luminosity,
$L_{\GW}(t)$, from Eq.~(\ref{e:L_heat}).
The observed flux depends on the disk parameters through $t_{\cool}(r)$.
The results are particularly simple in two limiting cases, namely when
$t_{\cool}$ is much smaller or much larger than the heating timescale, for
which $\Delta F(t_{\ret}, r)/F_{\rm disk}(r)=\dot e_{\rm heat}/\dot e_{\rm
disk}\times \{1 \text{~or~} \Delta
t_{\GW}/t_{\cool}\exp(-t_{\ret}/t_{\cool})\}$, respectively.
Substituting Eq.~(\ref{e:e_heat/e_disk}) in these simple cases gives
\begin{equation}\label{e:DF}
 \frac{\Delta F(t_{\ret}, r)}{F_{\rm disk}(r)} \approx \left\{
\begin{array}{ll}
 \frac{32}{9} \Iota(\theta) r_3 L_{-3}^{\GW}(t_{\ret}) & \text{~~if~}
 \Delta t_{\GW}\gg t_{\cool}\\[1.5ex] \frac{16}{9} \kappa \Iota(\theta)
 r/(ct_{\cool}) & \text{~~if~} \Delta t_{\GW}\ll t_{\cool}
\end{array}
\right..
\end{equation}
\begin{figure}[tb]
\centering
\mbox{\includegraphics[height=2.5in]{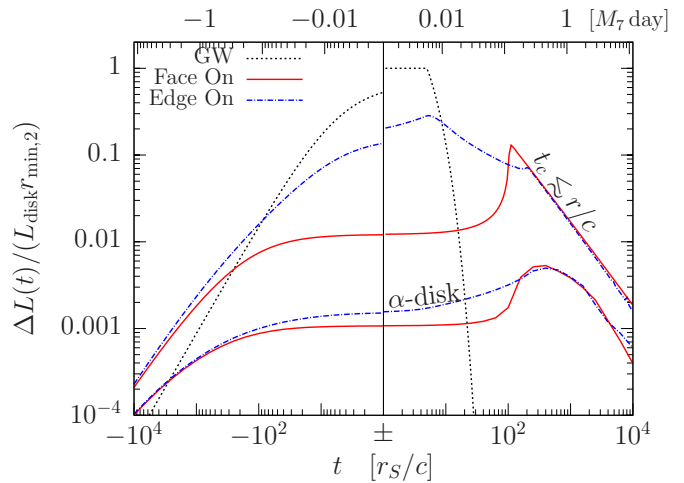}}
\caption{\label{f:lightcurve} The excess luminosity curve before ($t<0$)
and after ($t>0$) the binary coalescence event, relative to the luminosity
of the punctured disk. The time axis is shown on a logarithmic scale at
both negative and positive values (in units of $r_S$ at the bottom or
`$M_7$ days' at the top). The solid and dashed curves correspond to cases
where the accretion disk is face on or edge on, respectively. The top curve
applies if the cooling time is shorter than the geometrical delay $r/c$ at
each radius, while the bottom curve show the case where disk cooling is
limited by photon diffusion.
The dotted curve depicts $L^{\GW}_{-3}(t)$ for reference.
The $y$-axis is normalized by the luminosity of the punctured disk for a cavity radius, $r_{\rm min}=
10^2 r_{\rm min,2} r_S.$}
\end{figure}
For example, substituting $L_{-3}^{\GW}=1$, $Y=1$, $\kappa=3\%$, and $r/(ct_{\cool})=0.1$,
we find that ${\Delta F(r)}/{F_{\rm disk}(r)}\sim 3.6 r_3$ and $0.005 $ in the
two cases, respectively. Depending on the binary orientation relative to
the disk, $Y$, the result can be a factor of $\sim 3$ larger or smaller
(see Eq.~\ref{e:eGW}). The net excess in the apparent luminosity of the
disk is
\begin{equation}\label{e:totallum}
 \Delta L(t) =  \frac{\cos\theta_{\rm obs}}{4\pi d_{L}^2}\int\limits_{r_{\min}}^{r_{\max}} \int\limits_0^{2\pi}  \Delta F(t'_{\rm ret},r)  \,r\D\phi_{\rm disk}\D r.
\end{equation}
Here, $t'_{\rm ret}=t-(r/{\rm c})(1-\sin\theta_{\rm obs}\cos \phi_{\rm
disk})$, where $\theta_{\rm obs}$ is the inclination of the disk relative
to the line of sight, $\phi_{\rm disk}$ is the azimuthal angle within the
disk, and $d_{L}(z)$ is the luminosity distance to the source at redshift
$z$.

We integrate Eq.~(\ref{e:heatingbal}) exactly for our fit to $L_{\GW}(t)$
and find the luminosity excess $\Delta L(t)$ using
Eq.~(\ref{e:totallum}). Figure~\ref{f:lightcurve} shows
the result relative to the punctured disk luminosity $L_{\rm disk}$.
We assume that the disk has an inner cutoff at $10^2 r_{\min,2}r_S$, and adopt
$(\q,\dot m,\alpha, b, M_7, Y)=(1, 0.1,0.3,0,1,2.5)$,
We consider two cases: {\it (i)} $t_{\cool}\ll r/{\rm c}$; and {\it (ii)} $t_{\cool}(r)$ for the
$\alpha$-disk model.  The dotted (blue) and solid (red) curves correspond
to $\theta_{\rm obs}=0$ and $\pi/2$, respectively.  The light--curve is
highly sensitive to $\theta_{\rm obs}$ due to the geometric (GW
travel-time) delay. If observed in the disk plane, the disk flare at all
radii is seen coincidently along the line--of--sight to the SMBHs, but with
a delay in other directions. However, if the disk is observed face--on,
then different annuli are seen coincidently and the geometric delay
increases with radius. In the latter case, the peak EM luminosity is
delayed by $t_{\cool}(r_{\min}) + (r_{\min}/{\rm c})\cos\theta_{\rm obs}$
relative to the GW peak.  The light curve has a characteristic shape, with
$\Delta L\propto |t|^{-5/4}$ in the inspiral phase, $\Delta L\propto
t^{-1}$ after the peak for a time $\sim t_{\rm cool}\sim 10 r_{\max}/{\rm
c}$, and an exponential decay at later times.
For unequal masses, $\Delta L$ is reduced by $\q^2$.

\paragraph*{Discussion.---}
Figure~\ref{f:lightcurve} implies that the excess luminosity of
a thin circumbinary disk peaks with a delay $\sim 10 M_7$ hours
relative to the peak in the GW burst. The magnitude of the excess
luminosity peak is $\Delta L/ L_{\rm disk}\sim (r_{\min}/10^{3}r_S)$ for
$t_{\cool}\ll r/c$, and a factor of $\sim 130 \dot m^{4/5} M_7^{1/5}$
smaller for more realistic $t_{\cool}(r)$ values.  This amounts to $\sim
(10^{-4}$--$10^{-3})\times L_{\Edd}$.  During the $t^{-1}$ decline of the
EM transient, the characteristic emission wavelength corresponding to the
surface temperature of the disk $\sim 3.6\times 10^{3} r_3^{-3/4}{\,\rm K}$
is in the infrared band $\lambda\sim 1.5 r_3^{3/4}\,\mu{\rm m}$, and
increases in proportion to $t^{3/4}$ as the GW propagates outwards.  Future
monitoring surveys, such as PAN-STARRS or LSST
\footnote{http://pan-starrs.ifa.hawaii.edu; http://www.lsst.org}, could
search for the expected transients \cite{Kocsis07}.

In radiatively-inefficient (geometrically-thick) accretion flows
\cite{ny94}, the heating rate is also given by a universal expression similar
to Eq.~(\ref{e:e_heat/e_disk}). However, due to the low radiation efficiency,
the resulting light curve is fainter and difficult to observe. We also find
that GW heating is not sufficient for reversing the
flow away from the BH.

Measurement of the GW heating effect would provide an indirect detection of
GWs with traditional electromagnetic observatories, and test the theory of
general relativity for the interaction of GWs with matter.  Identification
of an unambiguous EM counterpart to a LISA measurement of GWs, would enable
to determine the source redshift and location as well as to constrain the
gaseous environments of merging SMBH binaries.

\bigskip
\acknowledgments
We thank Z. Haiman, K. Menou, R. Narayan, G. Rybicki, and S. Shapiro for
helpful comments. BK acknowledges support from OTKA Grant 68228 and Pol\'anyi Program of the Hungarian National Office for Research and Technology
(NKTH).

\bibliography{ms}
\end{document}